\documentclass[conference]{IEEEtran}
\IEEEoverridecommandlockouts
\usepackage{cite}
\usepackage{amsmath,amssymb,amsfonts}
\usepackage{algorithmic}
\usepackage{graphicx}
\usepackage{textcomp}
\usepackage{xcolor}
\usepackage{array}
\usepackage{tikz}

\newcommand\copyrighttext{%
  \footnotesize \textcopyright 2026 IEEE. Personal use of this material is permitted. Permission from IEEE must be obtained for all other uses, in any current or future media, including reprinting/republishing this material for advertising or promotional purposes, creating new collective works, for resale or redistribution to servers or lists, or reuse of any copyrighted component of this work in other works.}
\newcommand\copyrightnotice{%
\begin{tikzpicture}[remember picture,overlay]
\node[anchor=south,yshift=10pt] at (current page.south) {\fbox{\parbox{\dimexpr\textwidth-\fboxsep-\fboxrule\relax}{\copyrighttext}}};
\end{tikzpicture}%
}

\def\BibTeX{{\rm B\kern-.05em{\sc i\kern-.025em b}\kern-.08em
    T\kern-.1667em\lower.7ex\hbox{E}\kern-.125emX}}

\newcommand{\slot}[1]{$\langle$\textit{{#1}}$\rangle$}
\newcommand{\prompt}[1]{%
  \vspace{1em}
  \hspace{0.03\linewidth}%
  \begin{minipage}{0.94\linewidth}%
    \hrule\vspace{0.5em}
    \setlength{\parindent}{0pt}%
    #1
    \vspace{0.5em}\hrule
  \end{minipage}
  \vspace{1em}
}

\begin{document}

\title{Customer Analysis and Text Generation for Small Retail Stores Using LLM-Generated Marketing Presence
\thanks{This work was supported in part by JSPS KAKENHI Grant Numbers JP24K03052 and JP22K18006.}}

\author{\IEEEauthorblockN{Shiori Nakamura}
\IEEEauthorblockA{\textit{Nagoya Institute of Technology}\\
Nagoya, Aichi, Japan \\
snakamura@ozlab.org}
\and
\IEEEauthorblockN{Masato Kikuchi}
\IEEEauthorblockA{\textit{Nagoya Institute of Technology}\\
Nagoya, Aichi, Japan \\
kikuchi@nitech.ac.jp}
\and
\IEEEauthorblockN{Tadachika Ozono}
\IEEEauthorblockA{\textit{Nagoya Institute of Technology}\\
Nagoya, Aichi, Japan \\
ozono@nitech.ac.jp}
}

\maketitle
\copyrightnotice{}

\begin{abstract}
Point of purchase (POP) materials can be created to assist non-experts by combining large language models (LLMs) with human insight. Persuasive POP texts require both customer understanding and expressive writing skills. However, LLM-generated texts often lack creative diversity, while human users may have limited experience in marketing and content creation. To address these complementary limitations, we propose a prototype system for small retail stores that enhances POP creation through human-AI collaboration. The system supports users in understanding target customers, generating draft POP texts, refining expressions, and evaluating candidates through simulated personas. Our experimental results show that this process significantly improves text quality: the average evaluation score increased by 2.37 points on a -3 to +3 scale compared to that created without system support.
\end{abstract}

\section{Introduction}
Point-of-purchase (POP) materials are critical in influencing consumer behavior~\cite{6}, particularly in small-scale retail settings such as apparel stores. A POP display typically includes product names, prices, and product explanations while its sales impact is known to be strongly influenced by the quality of its textual content. Therefore, an effective POP depends on attractive and targeted language to appeal to potential customers~\cite{7}. In addition to product knowledge, POP creators must understand the target customers, possess marketing expertise, and have writing skills that effectively convey the product's appeal~\cite{8,9}.

In practice, POP displays in small apparel retail stores are often created by non-specialists, as these stores rarely employ staff with dedicated expertise in POP design. Consequently, producing persuasive POP texts can be challenging for such individuals, requiring considerable time, effort, and experience.

To address this issue, we propose a \textit{POP Text Creation Support System} designed for small apparel retailers. The system decomposes the POP creation process into four steps, each realized as a dedicated module: 
(1) Profile Builder (PB), which assists the user in understanding the product and identifying the target customer; 
(2) Draft Generator (DG), which generates an initial POP text draft; 
(3) Style Rephraser (SR), which creates diverse rephrasings to enhance expressiveness and variety; and 
(4) Persona Evaluator (PE), which evaluates the generated texts from multiple persona-based perspectives.
Each module, implemented using a large language model (LLM), is integrated into a prototype system that enhances both customer understanding and text generation. 

The contributions of this study are twofold. First, we design and implement a modular POP Text Creation Support System that leverages LLMs to assist non-expert users in identifying target customers, generating persuasive texts, and refining expressions through human-AI collaboration.
Second, we conducted an evaluation experiment in which non-expert evaluators rated the quality of POP texts, demonstrating that our system significantly improves perceived text quality, with an average score increase of 2.37 points on a -3 to +3 scale compared to manual creation without system support.

\section{Related Work}
This section describes existing studies on prompt engineering to support text creation and a method for analyzing target customers using LLMs.

\subsection{Consumers and Products Matching}

Estimating which products consumers are likely to purchase by analyzing customer behavior data, such as reviews and purchase histories, is a well-established approach to connecting customers with products.
Chang et al.~\cite{1} developed and evaluated a conversational inference system that recommends optimal products by eliciting customer preferences.
In a simulation, the system achieved an approximately 70\% success rate in correctly identifying product features from a single query.

Small apparel retail stores often cannot accumulate sufficient customer data, such as reviews and purchase histories, to apply traditional data-driven recommendation methods.
In this context, we draw on a key finding from the study by Chang et al., which demonstrated that LLMs can infer user preferences from product features even without extensive prior data.
Building on this insight, we hypothesize that LLMs may also be effective in inferring potential customer segments based solely on product information, a reverse approach compared to traditional recommendation methods.

\subsection{Idea Generation and Extension Support}

An effective strategy for generating high-quality ideas is to produce many options and then select the most appropriate ones from the pool~\cite{10}.
However, manually generating many ideas poses a significant challenge. Consequently, researchers and practitioners have increasingly turned to LLMs to support and enhance the process of idea generation.
For example, Meincke et al.~\cite{2} conducted a study showing that GPT-4 can produce higher-quality ideas than humans, although its outputs tend to lack originality and exhibit considerable content overlap.
They compared 200 ideas generated by university students with 100 ideas produced by GPT-4 using zero-shot and few-shot prompts.
Their analysis revealed that GPT-4's ideas scored higher on average regarding purchase intention.
However, they also noted that the AI-generated ideas were generally less unique than those created by humans.
Furthermore, some studies used LLMs to generate ideas, expand, and evaluate them.
For example, the writing support systems described in \cite{3} and \cite{4}, proposed additional ideas based on those presented by the user, made stylistic changes, optimized the document structure, and offered feedback and suggestions for improvement.
Although these studies showed that these stylistic changes and suggestions improve writing quality, generating truly original additional ideas still poses a challenge.

Based on the above, LLMs may not be ideal for generating highly original ideas. However, they excel at tasks such as sentence refinement and style transformation.
Therefore, instead of relying on fully automated text generation, our system leverages LLMs to assist creators in composing POP text.
In particular, functions such as rephrasing are effective in enhancing the expressiveness of POP displays.

\subsection{Subjective Evaluation Methods for Ideas}

Subjective evaluations---such as interest, engagement, and utility---produced by conventional automated evaluation methods using LLMs often suffer from inconsistency.
Wu et al.~\cite{5} proposed an ensemble evaluation method based on multi-role prompts as a solution.
The Wu method assigns different roles to LLMs (e.g., critic, general reader), allowing them to assess texts from both objective (e.g., grammar, consistency) and subjective (e.g., engagement, utility) perspectives.
This approach enables multi-dimensional evaluation by incorporating more diverse viewpoints and is highly consistent with human evaluation.

To enhance the POP text's appeal, it is desirable to incorporate the perspectives of the customers who are the target audience.
We designed a module to obtain multidimensional feedback by dynamically generating multiple personas that represent various customer segments, allowing POP creators to select the text more efficiently that best resonates with their intended audience.

\section{POP Text Creation Support System}

In this study, the task of POP creation is decomposed into the following four subtasks:  
(1) understanding the product and identifying the target customer,  
(2) generating an initial POP text draft,  
(3) refining the draft through diverse rephrasings, and  
(4) evaluating the generated texts from multiple perspectives.  

We developed a prototype system that supports each of these subtasks using a modular architecture. The overall system, referred to as the \textit{POP Text Creation Support System}, consists of four components:  
Profile Builder (PB), which assists the user in understanding the product and identifying the target customer;  
Draft Generator (DG), which produces an initial POP text draft;  
Style Rephraser (SR), which generates multiple rephrased variants to enhance expressiveness and diversity; and  
Persona Evaluator (PE), which evaluates the drafts from multiple persona-based perspectives.  

Each component is implemented using a LLM integrated into a unified system that supports both customer understanding and POP text generation.

\begin{figure}[tb]
  \centerline
  {
    \includegraphics[clip,width=7.5cm]{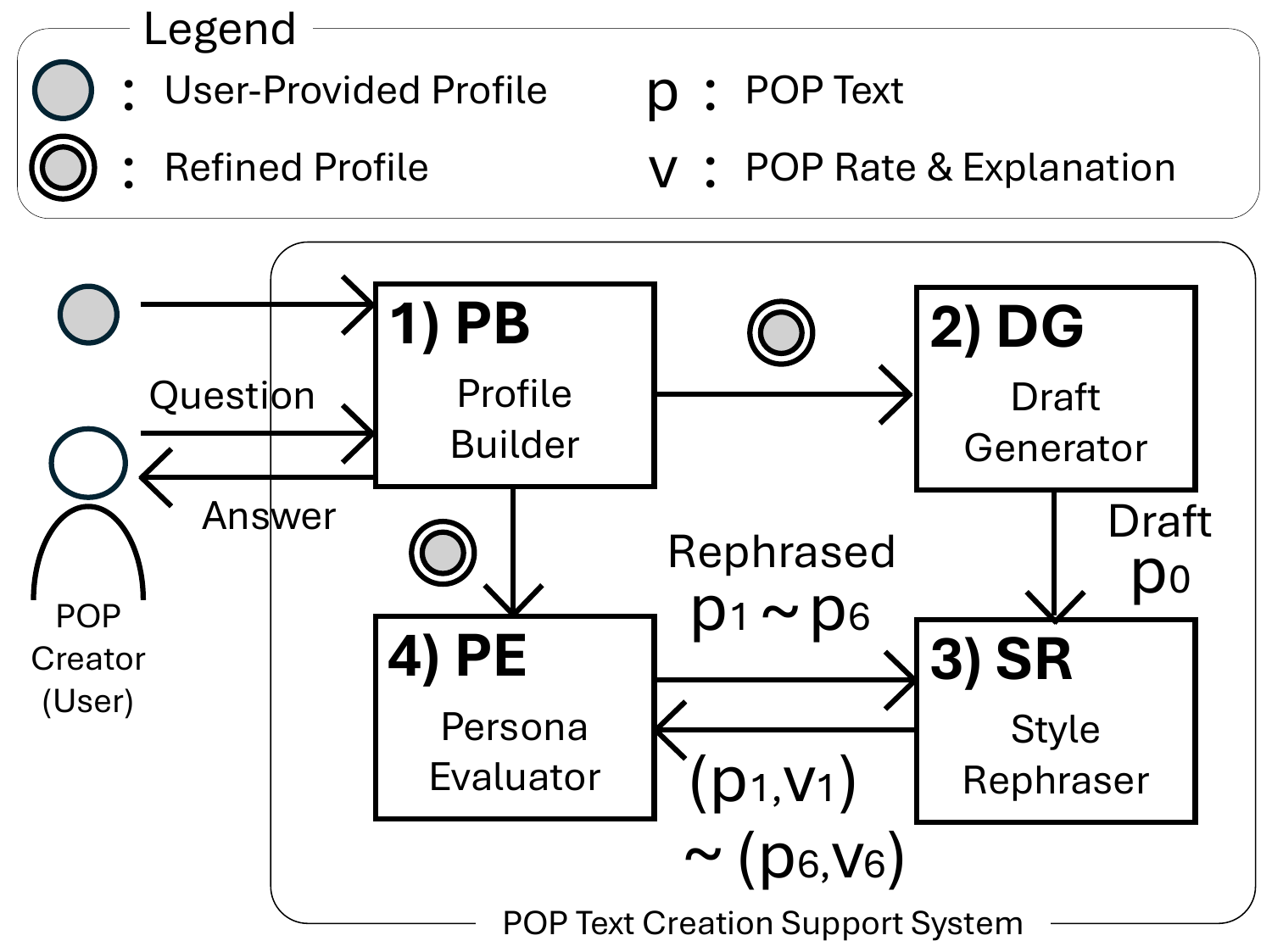}
  }
    \caption{System architecture}
    \label{fig:figure_1}
\end{figure}

\textbf{Profile Builder (PB)} obtains input from a user (i.e., a POP creator) regarding the target customers and product information (Fig.~\ref{fig:figure_1}: Target Customer and Product: User-Provided Profiles (User-Provided Profiles)) and asks the user a question about this information.
PB then obtains the answer from the user.
By repeating this question-and-answer process, PB augments the user's understanding of the customer.
Consequently, PB outputs the specific customer and product information (Fig.~\ref{fig:figure_1}: Target Customer and Product: Refined Profiles (Refined Profiles)).
Whenever a user answers a question, it updates the Refined Profiles and passes this information to DG and PE.
DG generates a POP text draft every time a user answers a question.
By referring to this draft, this system assists the user in deciding whether the target customer should be analyzed further through additional questions.

\textbf{Draft Generator (DG)} generates a POP text draft p$_0$ from the Refined Profiles.
Then, this draft is passed to SR.
Although it is intended to appeal to the target customer, there is still room for improvement, such as tailoring the text to the specific characteristics of the store.
This study attempted an approach in which a POP text draft is first created using the Refined Profiles, and is then improved by SR to fit other information, such as store characteristics.

\textbf{Style Rephraser (SR)} generates multiple candidates for POP text by rephrasing the draft p$_0$ received from DG.
Here, for example, rephrased patterns such as ``emphasize price'' and ``emphasize trend'' are pre-defined.
Both p$_0$ and the POP texts generated by SR (p$_1$ to p$_6$ in the figure), are subject to rephrasing.
This makes it possible to repeat rephrasing until a POP text is generated that satisfies the user.
The POP texts generated by SR (p$_1$ to p$_6$ in the figure) are passed to PE.

\textbf{Persona Evaluator (PE)} evaluates the goodness of the POPs obtained as input.
To generate evaluations from multiple perspectives, PE uses various personas (e.g., a fashion-conscious 20-something woman) to reflect the target customer base.
Each persona outputs POPs evaluation values and reasons for the evaluation, taking into account both the POPs and the Refined Profiles received from PB.

For this research, we used GPT-4o mini provided by OpenAI. 
The model was accessed via the OpenAI API using the official Python SDK (openai library), and prompts were sent in a conversational format using the chat.completions.create() method. 
All model parameters, including temperature and top-p, were set to their default values as provided by OpenAI.
The following sections describe the implementation of each module.

\subsection{Profile Builder (PB)}

\begin{figure}[tb]
  \centerline
  {
    \includegraphics[clip,width=7.5cm]{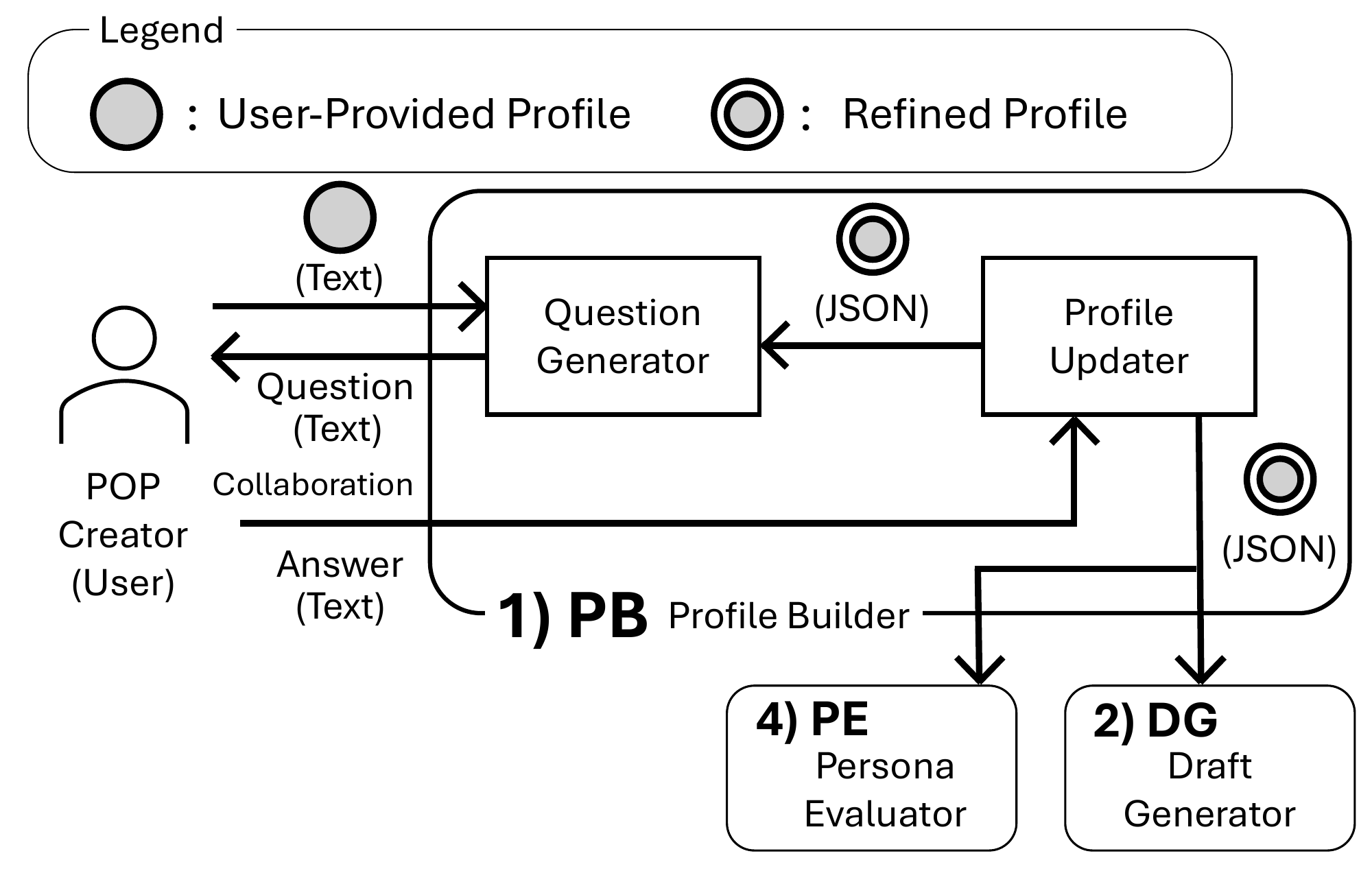}
  }
    \caption{ The Profile Builder System architecture}
    \label{fig:figure_2}
\end{figure}

The PB module is designed to assist the user (i.e., the POP creator) gain a deeper understanding of the attributes of the product and its target customer.
Fig.~\ref{fig:figure_2} shows the PB system architecture.
First, PB uses an LLM to generate Yes/No questions based on the product information and the attributes of the target customer---such as gender and age---provided by the user (Fig.~\ref{fig:figure_2}: User-Provided Profiles).
When the user answers a question, the target customer and product information is compiled based on the response to the question (Fig.~\ref{fig:figure_2}: Refined Profiles is created from the Profile Updater).
This improved information is then passed to DG, PE, and the Question Generator.
By using the Refined Profiles, this function can generate in-depth questions based on previous answers.

Fig.~\ref{fig:prompt_1} shows the prompt used by PB.
In this figure, slots that must be completed are suggested by \slot{}, and the type of information that should be included is shown.
The required information includes details about the product, the target customer, and past questions and answers, with instructions to generate Yes/No questions based on this information.

\subsection{Draft Generator (DG) and Style Rephraser (SR)}

DG and SR are modules for generating POP text drafts.
First, DG generates an initial POP text draft. It receives the Refined Profiles from PB, which contains information about the product, target customer, and past interactions. Using this information, DG employs an LLM to generate a POP text draft, p$_0$, including a catchphrase and product explanation tailored to the customer. Next, SR rephrases p$_0$ into multiple variations, creating several new POP text drafts with various expressions.

\begin{figure}[tb]
  \begin{flushleft}
    \prompt{
    You are an assistant helping the clerk perform ``proper customer segmentation of merchandise''.
    To do this, think of a question. 
    This question asks the clerk, ``Is this product for this type of person?''
    Present another question based on the product information and the previous question and its answer.
    Output the question text and the reason for proposing the question.
    The product is for 	\slot{gender and age of the target customer}.
    Product explanation: 	\slot{information about the product obtained by free input} The answers so far are as follows.
    $\langle\textit{list the text of the question and the reason why LLM}$
    $\textit{choses this question and the answer}\rangle$
    }
    \caption{Prompt for the Profile Builder (PB)}
    \label{fig:prompt_1}
  \end{flushleft}
\end{figure}

The POP text drafts generated by DG are then rephrased into various expressions by SR.
The purpose of DR is to present a variety of ideas to the user (i.e., the POP creator) to encourage the comparison and combination of ideas and support the creation of more appealing text.
Specifically, SR generates multiple new drafts by rephrasing the original POP text draft using LLM.

\begin{figure}[tb]
  \centerline
  {
    \includegraphics[clip,width=6.75cm]{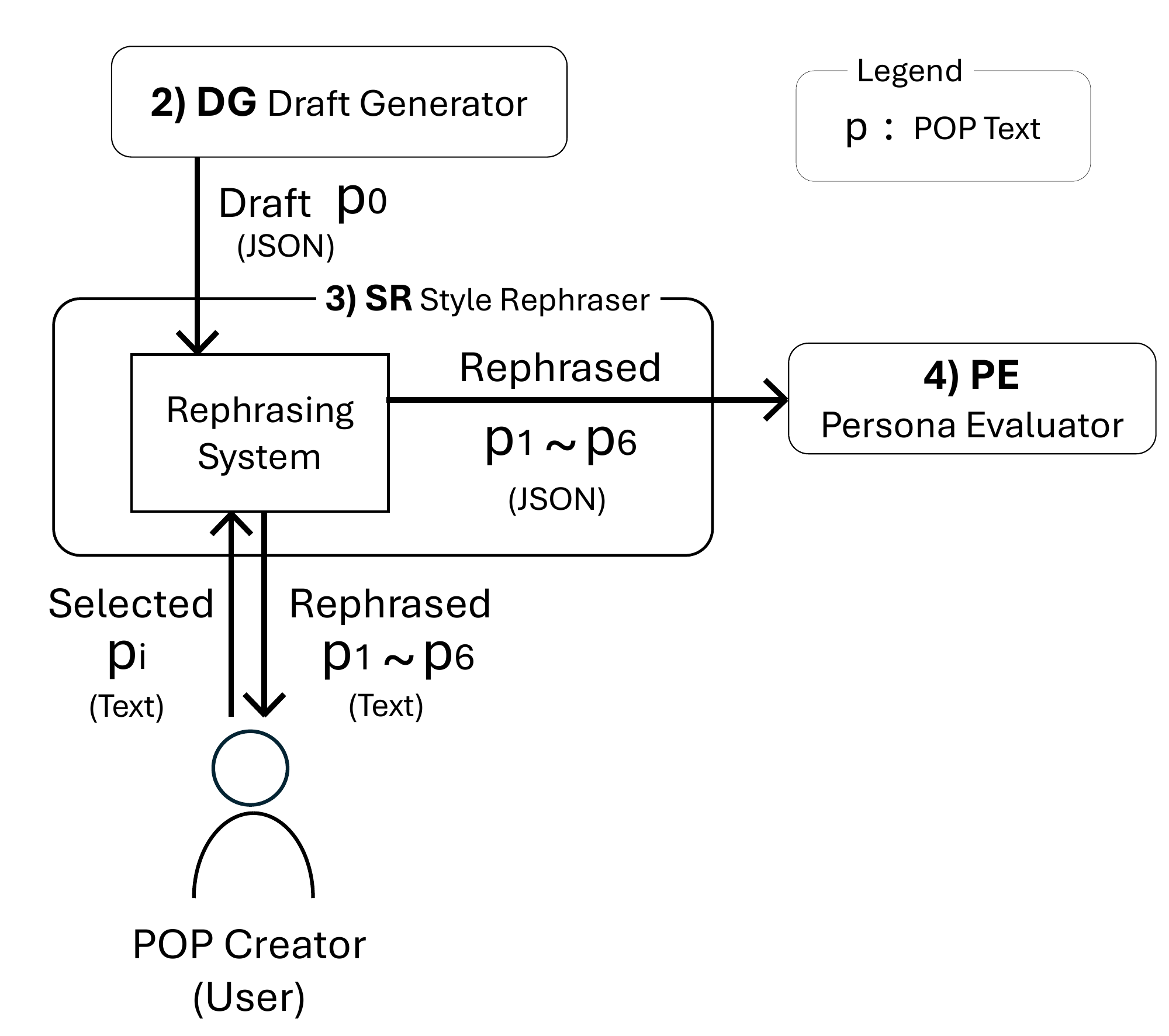}
  }
    \caption{System architecture of the Style Rephraser (SR)}
    \label{fig:figure_5}
\end{figure}

Fig.~\ref{fig:figure_5} shows the SR system architecture.
First, SR receives a POP text draft p$_0$ from DG.
Then, based on this draft, SR generates six new drafts p$_1$ to p$_6$ using LLM (Fig.~\ref{fig:figure_5}: Rephrasing System).
These six rephrasing patterns are derived from the factors that influence the purchase of clothing (appearance preference/suitability, fashionability, practicality/economy, quality/traditionality/reliability, gaining others' approval, and combination).
In addition, the POP text draft generated by DG, the POP text draft generated by SR, and the rephrasing can be repeated by the user by selecting the POP text draft p$_i$ that he/she wants to use as the source of the text.
The POP text generated by SR is then passed to PE.

Fig.~\ref{fig:prompt_3} shows the prompt used by the Rephrasing System.
In this figure, the slots that need to be filled are suggested by \slot{}, and the type of information that should be included in the slots is shown.
The prompt explicitly asks for a catchphrase, a product explanation, and a purchase motive on which it can focus.
The purchase motive slot can be set to appearance preference/suitability, fashionability, practicality/economy, quality/traditionality/reliability, gaining others' approval, and combination.
The expressions are then adjusted depending on which purchasing motive is emphasized. 
When concrete instructions are provided, they create a fixed framework for the generated text drafts. Consequently, the drafts tended to become biased toward similar patterns. 
Conversely, abstract instructions allow for greater diversity in the generated text drafts.
Therefore, we retained the purchase motives in the prompt as abstract instructions to encourage a variety of expressions.

\begin{figure}[tb]
    \prompt{
    	\slot{original catchphrase}\\
    	\slot{original product explanation}\\
    	Rephrase this into a sentence focusing on the \slot{purchase motivation}.
    	The catchphrase should be approximately 10 characters long and the product explanation should contain approximately 50 characters, counting the number of characters.
    
    	The slot for the purchase motivation can be one of the following: appearance preference/suitability, fashionability, practicality/economy, quality/traditionality/reliability, gaining others' approval, and combination.
    }
    \caption{Prompt for Style Rephraser (SR)}
    \label{fig:prompt_3}
\end{figure}

\subsection{Persona Evaluator (PE)}

\begin{figure}[tb]
  \centerline
  {
    \includegraphics[clip,width=7.5cm]{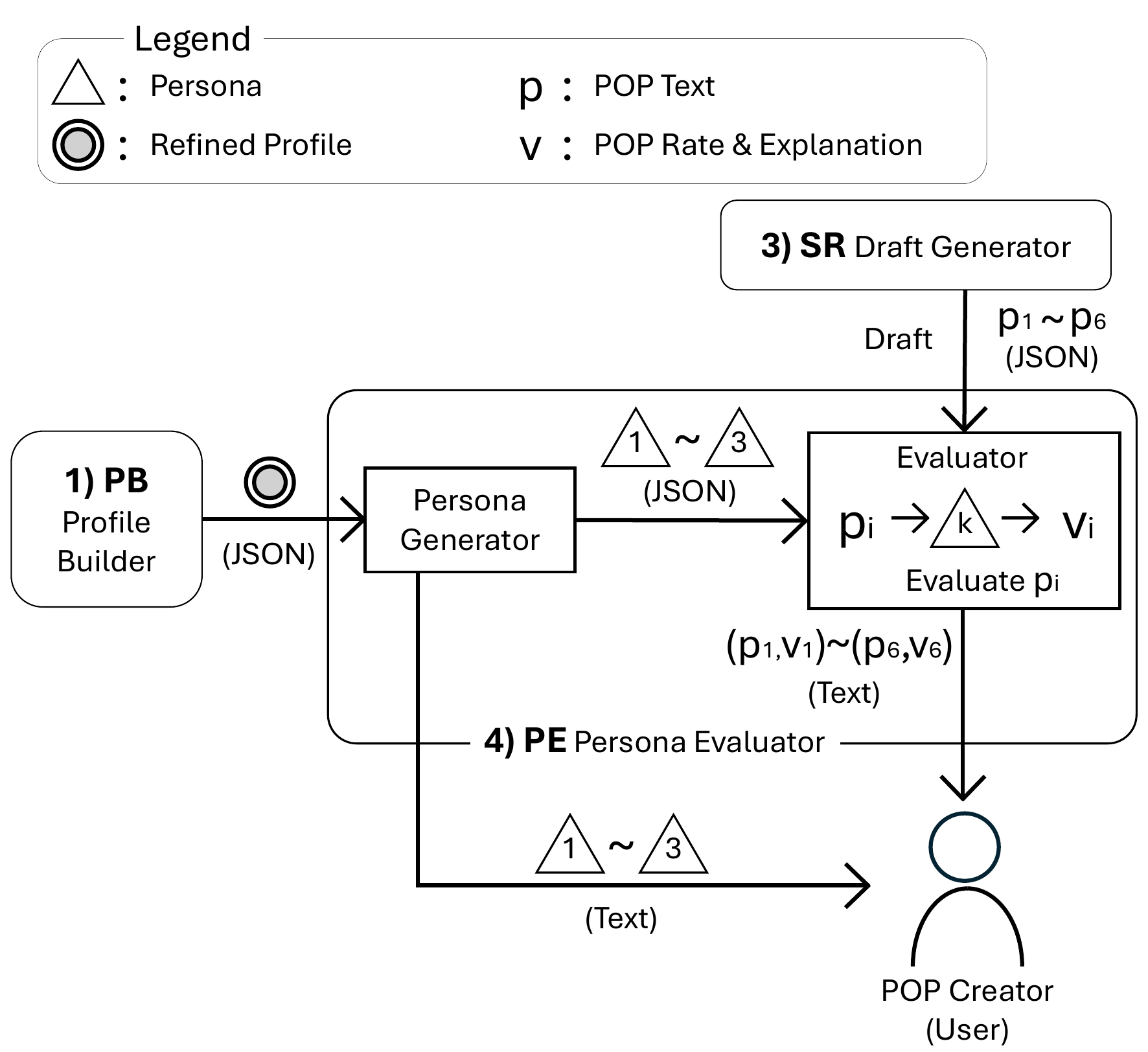}
  }
    \caption{Persona Evaluator (PE)}
    \label{fig:figure_7}
\end{figure}

PE is a module that supports the selection and creation of appropriate POP text drafts by presenting the evaluation of POP text drafts to the user.
The purpose of PE is to guide the convergent process of POP texts, originally generated in a divergent manner by SR, by providing evaluations from various viewpoints. 
As Fig. \ref{fig:figure_7} shows, PE comprises two sub-modules.
The Persona Generator generates three types of personas that correspond to the target customers.
The Evaluator presents evaluation values and justifications by having these personas assess the POP texts.

Personas are generated dynamically.
Each time the user answers a PB question personas are also regenerated based on the updated Refined Profiles.
These three personas are passed to the Evaluator, which then uses the LLM to evaluate the six POP texts (p$_1$ to p$_6$) generated by SR.
As such, each POP text is evaluated by three different personas, producing 18 evaluations in total. The pairs of POP texts and their evaluations, (p$_1$, v$_1$) through (p$_6$, v$_6$), are presented to the user.

Since personas must be tailored to each application domain, the prompts in this case are designed to generate personas that specialize in evaluating apparel POP texts.
Fig. \ref{fig:prompt_4} shows the prompt used by the Persona Generator.
In this function, the slots that must be filled are suggested by \slot{}, and the type of information that should be included in the slots is shown.
This prompt explicitly instructs the LLM to generate personas using the Refined Profiles.
The personas generated by this function have the following attributes: age, occupation (including housewife), family structure, lifestyle, three clothing needs, and three points that make the clothing attractive.
To evaluate the appeal of the text accurately, we decided to generate personas whose attributes align with purchasing motives.
 
Conversely, the personas generated by this function are expected to be useful for evaluating POP text and for understanding what kind of person the target customer's profile represents.
Importantly, these personas must include attributes that support both accurate text evaluation and a deeper comprehension of the target customer's characteristics.
For example, in the case of the practicality/economy purchasing motive, a persona's occupation and family structure are relevant. Therefore, we generated occupation and family structure attributes for personas whose purchasing motive is practicality/economy.

\begin{figure}[tb]
  \prompt{
  You are an assistant required to create three appropriate personas.
  You will be asked to evaluate the POP text according to the personas you have created.
  To do this, the target customers for the product have been narrowed down by means of questions.
  Based on this information, create personas (age, occupation (including housewife), family structure, lifestyle, 3 clothing needs, and 3 points that make the clothing attractive).
  Output only the personas.
  This is a product for \slot{target age and gender}
  Product explanation: \slot{product description}
  The previous answers are:.
  \slot{question, justification, and response history}
  }
  \caption{Prompt for the Persona Generator in Fig.\,\ref{fig:figure_7}}
  \label{fig:prompt_4}
\end{figure}

Fig.~\ref{fig:prompt_5} shows the prompt used by the Evaluator.
This figure shows the slots to be filled suggested by \slot{}, and the type of information that should be included in the slots.
In this prompt, the persona information and POP information are entered into the slots.
This instructs the LLM to rate the POP on a 10-point scale according to the persona's perspective, and briefly state the reason for the rating. 
By generating multiple personas and evaluating POP text based on each persona's viewpoints, subjective evaluations from various perspectives can be generated.
Providing a numerical rating offers a more convenient means of comparing different text drafts than relying solely on textual evaluation.

\begin{figure}[tb]
\prompt{
    You are an assistant who evaluates POP text based on three personas.
    Please rate each POP text on a 10-point scale based on each persona and provide a single reason for your rating.
    Do not output any other sentences.
    Persona: \slot{persona information}. POP texts: \slot{POP texts}.
}
    \caption{Prompt for Personal Evaluator (PE)}
    \label{fig:prompt_5}
\end{figure}

\section{Evaluation Experiment}
In the evaluation experiment, we evaluated PB, DG, SR, and PE.
The presence and absence of the above modules were varied, and changes in the quality of the POP texts were evaluated.
The evaluation of POP text quality was done manually.

\subsection{Experimental Procedure}

First, we describe the five types of comparison targets.
No support: creating POP texts without using this system, 
Analysis only: creating POP texts using only PB, 
Draft + Edit: creation of POP text drafts using PB and DG, followed by manual editing of POP text drafts without using this system,
All functions (manual selection): creating POP texts using this system. Manually editing POP text drafts and manually selecting a POP text draft by referring to the output of SR and PE.
All functions (automatic selection): creating POP texts using this system. Manually editing POP text drafts and selecting a POP text draft by PE as the best POP text.
In this experiment, we did not use the LLM-generated POP text drafts as-is but instead used a manually edited POP text draft. 
This approach was taken because the system is designed for scenarios where LLM-generated drafts serve as a reference and are manually refined.
Therefore, in this experiment, the POP text drafts generated by LLM were edited manually, and improvements and evaluations were done repeatedly.

The following explains the method for evaluating the POP text.
First, the evaluators were shown the POP texts created by the two different methods.
Next, the evaluators answered which POP text was better.
For example, for the comparison of POPs p$_1$ and p$_2$, if p$_1$ was better than p$_2$, an integer score between 1 and 3 was assigned to p$_1$.
The scores 1 to 3 represented the degree of goodness (with 3 representing the greatest degree).
In this case, p$_2$ was assigned the score of p$_1$ multiplied by -1.
For example, if p$_1$ had a score of 3, then p$_2$ had a score of -3.
Conversely, if p$_2$ was better, the signs of p$_1$ and p$_2$ were reversed.

For this experiment, one creator produced the POP text, and three evaluators---all non-specialists in POP creation---assessed it.
The creator, who had some POP creation experience but was not an expert, first asked the evaluators to select two clothing items they were interested in purchasing. 
For one selected product, the creator generated five types of POP texts using the five methods mentioned above. 
The evaluators then answered five different POP texts about the product they had selected, based on their own sense of value.

\subsection{Experimental Results}

\begin{figure}[tb]
  \centerline
  {
    \includegraphics[clip,width=7.5cm]{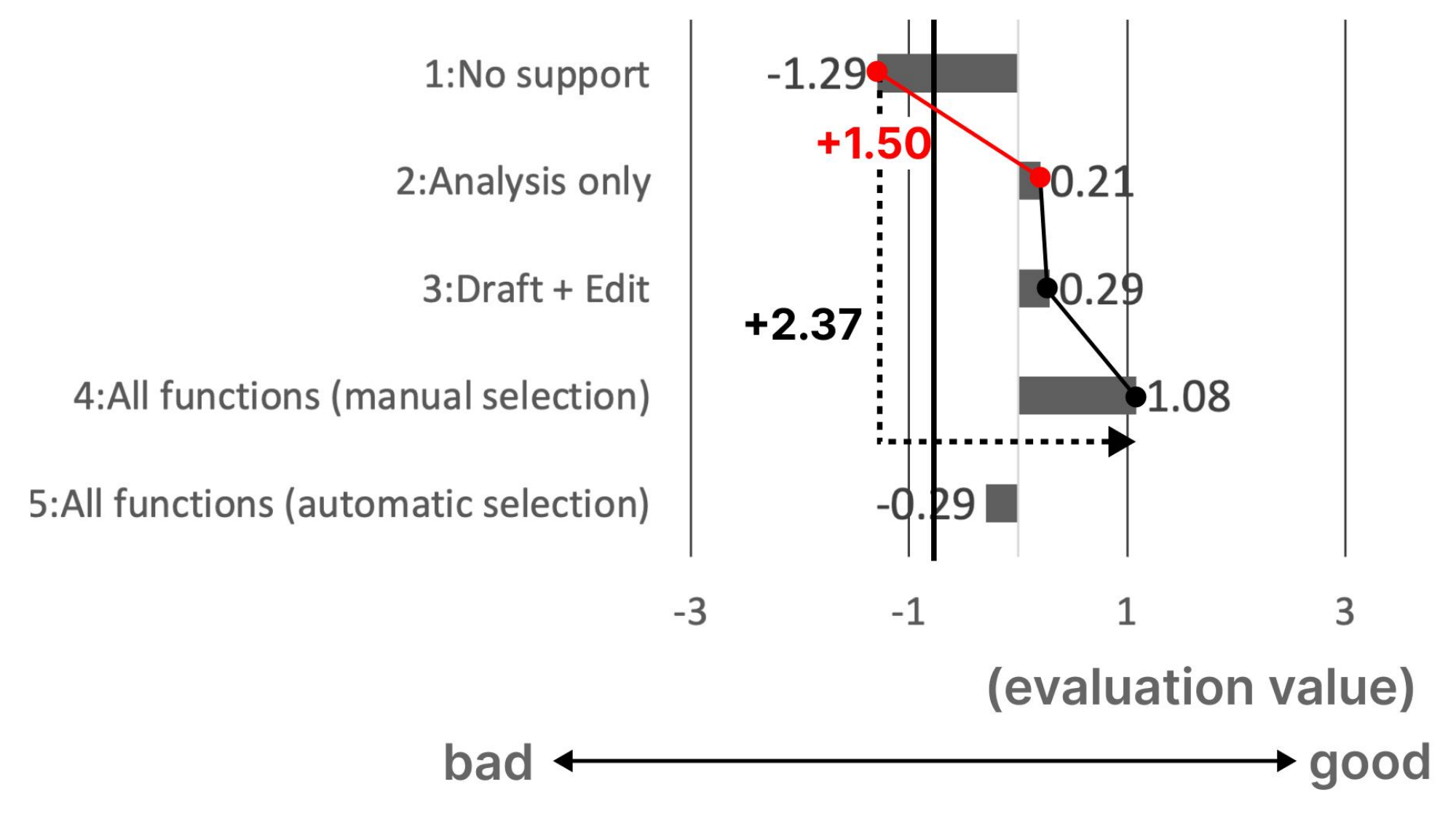}
  }
    \caption{The average evaluation score}
    \label{fig:figure_10}
\end{figure}

Fig. \ref{fig:figure_10} shows the results of the experiment.
From top to bottom, the results show the average score of the evaluation (-3 to +3) given to the POP texts created by methods 1 to 5.
The higher the evaluation score, the better the creation method.
As shown in this figure, 4: All functions (manual selection) is best.
Conversely, 1: No support is worst.
The average of the evaluation scores improved by 2.37 from 1 to 4.
The lower evaluation score for 5: All functions (automatic selection) compared to 4: All functions (manual selection) may be due to PE's output being useful as a reference but not reliable enough to determine the final selection on its own.
These results suggest the effectiveness of this system's support module.

The effectiveness of each module---No support, Analysis only, Draft + Edit, All functions (manual selection), and All functions (automatic selection)---was analyzed in more detail. 
First, comparing ``2: Analysis only'' and ``1: No support,'' ``2: Analysis only'' was judged to be preferable (higher evaluation score) to ``1: No support'' in 83.3\% of the POP texts.
This means that PB is effective even by itself in supporting the creation of attractive POP texts.
However, when comparing the POP texts created by ``4: All functions (manual selection)'' and ``5: All functions (automatic selection),'' the POP text created by ``4: All functions (manual selection)'' was judged to be better in 83.3\% of the comparisons.
Therefore, it can be said that the evaluations generated by PE should not be trusted blindly.

\section{Discussion}

The characteristics and problems of each system module were examined, along with the potential directions for improvement.
Persona Generator exhibits a problem, when personas were generated based on the target customers provided by PB, information not explicitly stated in the original questions was sometimes introduced as a feature of the persona.
For example, when the question was asked to narrow down the target customer group to ``a woman in her 50s who likes casual clothes and enjoys outdoor activities'', the generated persona may include the trait ``prefers environmentally friendly products''.
The evaluation of the proposed text focused mainly on environmental friendliness, and the original attributes of the expected customer were being neglected.
To increase the persona-based evaluation's reliability, the consistency between the persona and the customer analysis results must be managed.

Next, as mentioned, when comparing the POP texts generated by ``1: No support'' and ``2: Analysis only'', the ``2: Analysis only'' method tended to receive better evaluations.
PB helps actualize the target customer and further clarifies the product's points of appeal. 
As a result, specific and persuasive POP texts were obtained.
Table \ref{table:example-ana} shows a comparison of POP texts created with the ``1: No support'' method and those created with the ``2: Analysis only'' method.
In this example, the POP texts created with the ``1: No support'' method were not restricted to any target.
Conversely, the POP texts created using the analysis were aimed at the ``adult casual lovers'' segment.
Furthermore, the expression ``easy to wear'' in the ``1: No support'' method changed to ``can be used in a variety of situations'' in the ``2: Analysis only'' text.
This clarified that ``easy to wear'' does not refer to comfort but to ease of coordination.

\begin{table}[tb]
\renewcommand{\arraystretch}{1.3}
  \caption{Example of POP text based on\\ ``1: No support'' and ``2: Analysis only''}
  \label{table:example-ana}
\centering
    \begin{tabular}{|p{1.9cm}|p{2.3cm}|p{3cm}|}
      \hline
        \multicolumn{1}{|c}{\textbf{Method}} & \multicolumn{1}{|c}{\textbf{POP Text}} & \multicolumn{1}{|c|}{\textbf{Product Description}} \\
        \hline
        \hline
        1: No support & You must have one of these pants! & The one centerline of these pants creates a beautiful silhouette! They are conveniently wide pants!\\
        \hline
        2: Analysis only & Recommended for adult casual lovers! & One centerline of these pants will make you look beautiful and mature! It is convenient for a variety of occasions!\\
        \hline
    \end{tabular}
\end{table}

\section{Conclusions}

To solve the problem of POP text creation in small apparel retail stores, we proposed and developed a system that supports the analysis of target customers and the created text.
An evaluation experiment has shown that the system can assist non-specialists in creating texts suitable for the target customers.
In particular, we confirmed the improvement of the system's power by using Profile Builder(PB) and Style Rephraser (SR).
Specifically, the evaluation score increased by 2.37 points compared to the case of No support in the POP texts evaluation evaluated in the range of -3 to +3.
This is sufficient considering the range of evaluation scores.
However, the Evaluator's persona-based evaluation must be improved.
Additionally, further assessments and improvements are needed to adapt the system for use in different application areas.

\bibliographystyle{unsrt}
\bibliography{reference}

@inproceedings{6,
    author="Y. Kosaka and H. Shiizuka",
    title="A Modeling and Systems Thinking Approach to Activity Rousing Consumer's Buying Motivation Focusing on ``Kansei Information'' in POP ADS at the Store",
    year="2010",
    pages="597--606",
    booktitle="Proc. KES IDT 2010",
}

@article{7,
    author = {C. S. Areni and R. Miller},
    title = {Sales effects of in-store radio advertising},
    journal = {Journal of Marketing Communications},
    volume = {18},
    number = {4},
    pages = {285--295},
    year = {2012},
}

@inproceedings{8,
  title={Creation of Effective Advertising in the Persuasion of Target Audience},
  author={M. Usman},
  year={2013},
  pages={1--6},
  journal={International Journal of Economics, Finance and Management},
}

@article{9,
    title = {Nontarget Markets and Viewer Distinctiveness: The Impact of Target Marketing on Advertising Attitudes},
    journal = {Journal of Consumer Psychology},
    volume = {9},
    number = {3},
    pages = {127--140},
    year = {2000},
    author = {J. L. Aaker and A. M. Brumbaugh and S. A. Grier},
}

@inproceedings{1,
    title = "Conversational Product Recommendation using LLM",
    author = "T. J. Chang and L. H. M. Lin, R. T. H. Tsai",
    year = "2024",
    pages = "340--343",
    booktitle = "Proc. ICEIB 2024"
}

@article{10,
    title={Does Quantity Generate Quality? Testing the Fundamental Principle of Brainstorming},
    volume={8},
    number={2},
    journal={The Spanish Journal of Psychology},
    author={A. M. Adánez},
    year={2005},
    pages={215--220}
}

@article{2,
  title={Ideas are Dimes a Dozen: Large Language Models for Idea Generation in Innovation},
  author={K. Girotra and L. Meincke and C. Terwiesch and K. T. Ulrich},
  journal={SSRN Electronic Journal},
  year={2023},
  pages = {1--36},
}

@inproceedings{3,
    author = {T. Chakrabarty and V. Padmakumar and F. Brahman and S. Muresan},
    title = {Creativity Support in the Age of Large Language Models: An Empirical Study Involving Professional Writers},
    year = {2024},
    booktitle = {Proc. C\&C 2024},
    pages = {132--155},
}

@inproceedings{4,
    author = {A. G\"{o}ldi and T. Wambsganss and S. P. Neshaei and R. Rietsche},
    title = {Intelligent Support Engages Writers Through Relevant Cognitive Processes},
    year = {2024},
    pages={1--12},
    booktitle = {Proc. CHI 2024},
}

@inproceedings{5,
    author = {N. Wu and M. Gong and L. Shou and S. Liang and D. Jiang},
    title = {Large Language Models are Diverse Role-Players forandnbsp;Summarization Evaluation},
    year = {2023},
    pages = {695--707},
    booktitle = {Proc. NLPCC 2023},
}

\end{document}